\documentclass[12pt]{article}
\usepackage{graphicx}
\usepackage{dcolumn}
\usepackage{bm}
\usepackage{graphics}
\usepackage{amsmath}
\usepackage{amssymb}
\usepackage{amscd}
\usepackage{afterpage}
\usepackage{float,times}
\usepackage{subfigure}
\usepackage{rotating}
\usepackage{multirow}
\usepackage{epsfig}
\usepackage{theorem}
\usepackage{moreverb}
\usepackage{euscript}
\usepackage{psfrag}

\begin{document}

\begin{center}
{\hbox to\hsize{\hfill April 2008 }}

\bigskip
\vspace{6\baselineskip}

{\Large \bf

Dyons with hidden electric charges  and \\ the long-range magnetic force\\
}
\bigskip

\bigskip

{\bf 
Archil Kobakhidze  \\}
\smallskip

{ \small \it School of Physics, The University of Melbourne, Victoria 3010, Australia \\ 
E-mail: archilk@unimelb.edu.au \\}

\bigskip

\vspace*{.5cm}

{\bf Abstract}\\
\end{center}
\noindent
{\small 
When gauge $U(1)$ is spontaneously broken,  associated electric charges are screened. We argue that corresponding magnetic charges, on the contrary, produce long-range force.  Some interesting consequences of this phenomenon are also discussed.
 }  

\bigskip

\bigskip

\bigskip



\paragraph{1.}

In many theoretical models ordinary quarks and leptons carry hypothetical extra charges and participate in extra gauge interactions. These charges and interactions, if really exist, must show up at small distances (high energies), and are screened  on larger distances where current experiments have found no evidence for their existence. This is consistently achieved through the spontaneous breaking of those extra gauge symmetries [here we consider $U(1)$ Abelian symmetries only] by the Higgs mechanism, providing the extra gauge bosons with sufficiently large masses. In this paper we want to argue that although electric charges in spontaneously broken gauge theories  are screened in the Higgs vacuum, the magnetic charges still produce an unscreened long-range forces. Thus the magnetic charges can be constrained from "fifth-force" type of experiments.  On the other hand, we will also argue that in the presence of such magnetic charges the Dirac charge quantization \cite{Dirac:1931kp}-\cite{Zwanziger:1969by} is extended for the screened electric charges. Then, taking into account constraints on magnetic charges, we come to the conclusion that screened electric charges must be extremely large, so that the gauge field - charge system is out of the Proca regime even at very low energies and laboratory strength magnetic fields. This implies that hidden charges can be identified in Aharonov-Bohm type of tabletop experiments. 

Let us be more specific and consider a model governed by the Lagrangian
\begin{equation}
{\cal L}=-\frac{1}{4}X_{\mu\nu}X^{\mu\nu}+|D_{\mu}\phi|-V(\phi)+...,
\label{1}
\end{equation} 
where $X_{\mu\nu}=\partial_{\mu}X_{\nu}-\partial_{\nu}X_{\mu}$ is the field strength of the $U_{X}(1)$ Abelian gauge field $X_{\mu}$; $\phi(x)$ is the complex scalar field with $U_X(1)$ charge $Q^{(e)}_{\phi}$, and thus $D_{\mu}\phi=\left(\partial_{\mu}+iQ^{(e)}_{\phi}\right)\phi$. We also assume that standard model fermions do also interact with $U(1)_X$ gauge boson. These interaction are not shown in (\ref{1}) explicitly. We assume the potential of the form
 \begin{equation}
 V(\phi)=\frac{\lambda}{4}\left(|\phi |^2-v^2 \right)^2~,
 \label{2}
 \end{equation}
which provides spontaneous breaking of $U(1)_X$ symmetry. If $Q^{(e)}_{\phi}=Ng_e$, where N is some integer, there is residual discrete symmetry survived the breaking, $U(1)_X\to Z_N$. Otherwise, the symmetry is broken completely, $U(1)_X\to 1$. In this so-called Higgs phase there are two mass scales: one is the Higgs mass $M_{H}=\sqrt{2\lambda}\langle |\phi| \rangle=\sqrt{2\lambda }v$ and another is the gauge boson mass $M_X=\sqrt{2}Q^{(e)}_{\phi}v$. If we assume that $M_{H}>>M_{X}$, than at low energies ($< M_{H}$) we can freeze the Higgs field and consider only the massive gauge boson (the system behaves like type II superconductor),
\begin{equation}
{\cal L}=-\frac{1}{4}X_{\mu\nu}X^{\mu\nu}+\frac{1}{2}M_X^2{\cal X}
_{\mu}{\cal X}
^{\mu}+...~,
\label{3}
\end{equation}  
where 
\begin{equation}
{\cal X}
_{\mu}=X_{\mu}+\frac{1}{Q^{(e)}_{\phi}}\partial_{\mu}f~,
\label{4}
\end{equation}
$f={\rm Im} \log \langle \phi \rangle$. 
Before moving on, let us recall important difference between the massive theory with the Higgs mechanism and pure Proca theory for massive vector field with explicitly broken gauge invariance. Indeed, the Lagrangian (\ref{4}) in the Higgs phase maintains full gauge invariance under the transformations: $\delta X_{\mu}=i\partial_{\mu}\alpha(x),~ \delta f=-iQ^{(e)}_{\phi}\alpha(x)$ (so that ${\cal X}
_{\mu}$ is gauge invariant vector potential,  $\delta {\cal X}
_{\mu}=0$).   This gauge invariance of (\ref{1}) and (\ref{3}) implies that a gauge-variant conserved  current $j_{\mu} $ ( e.g., $j_{\mu}=\bar\psi\gamma_{\mu}\psi$, where $\psi$ is some fermionic field), have to couple to $X_{\mu}$ and not to ${\cal X}
_{\mu}$, 
\begin{equation}
{\cal L}_{\rm int}=Q_{\psi}^{(e)}j_{\mu}X^{\mu}~,
\label{5}
\end{equation}  
where $Q_{\psi}$ is an $U(1)_X$ charge. Now let us assume that aside the electric charges, we have magnetic charges as well. The equation of motions read:
\begin{eqnarray}
\partial_{\mu}F^{\mu\nu}=Q_{\psi}^{(e)}j^{\nu}-M_X^2{\cal X}^{\nu}~, \\ 
\partial_{\mu}\tilde F^{\mu\nu}=Q_{\psi}^{(m)}\tilde j^{\nu}~, 
\label{6}
\end{eqnarray} 
where $\tilde F_{\mu\nu}=\frac{1}{2}\epsilon_{\mu\nu\rho\sigma}F^{\rho\sigma}$ is the dual field strength, and $\tilde j_{\mu}$ is the dual magnetic current  with $Q^{(m)}_{\psi}$  being a magnetic charge. One can observe immediately that the familiar electric-magnetic duality of massless electrodynamics is broken in the massive case. Namely, while the mass term enters in the electric Gauss law (6), no mass term is involved in the magnetic Gauss law (7). Consequently, the electric charge density $Q_{\psi}^{(e)}j^{0}$ is screened by a screening charge density $\rho_{\rm screen}=-M_X^2{\cal X}^0$, so that the electric field falls off exponentially at distances $r>1/M_X$. For point-like charge  $j^{0}=\delta^{3}(\vec x)$ and   we obtain  
\begin{equation}
\vec E=\frac{Q_{\psi}^{(e)}}{4\pi r^3}e^{-M_Xr}\vec r~.
\label{7}
\end{equation}  
On the contrary, the magnetic Gauss law (7) is the same as in the massless case, and one may  suspect that the magnetic field must be long-range. To see that this is indeed the case we consider eqs. (6) and (7) in the static limit with point-like magnetic charge only. Then we have, 
\begin{eqnarray}
\vec \nabla \cdot \vec B=Q_{\psi}^{(m)} \delta^{3}(\vec x)~, \\
\vec \nabla \times \vec B=-m^2\vec {\cal X}~,
\label{8}
\end{eqnarray} 
where $\vec B=\vec \nabla \times \vec X$. It is evident that the above equations are solved by the singular Dirac vector potential (the gauge is taken where the singular Dirac string is located along the negative $z$-axis )
\begin{equation}
X_{r}=X_{\theta}=0~,~~X_{\phi}=\frac{Q_{\psi}^{(m)}}{r}\frac{1-\cos\theta}{\sin\theta}~,
\label{10}
\end{equation}
provided,
\begin{equation}
\vec {\cal X}=0 \to \vec X+\frac{1}{Q^{(e)}_{\phi}}\vec \nabla f=0~.
\label{11}
\end{equation}
Actually, the later relation  solves the equation of motion for $f$, $\Box f=-Q^{(e)}_{\phi}\partial_{\mu}X^{\mu}$. Therefore, (\ref{10}) and (\ref{11}) indeed form the legitimate solution of the equations of motion. Thus we came to the conclusion that magnetic field generated by the point-like magnetic charge in the massive case coincides exactly with the long-range magnetic field of the Dirac monopole \cite{Dirac:1931kp}, 
\begin{equation}
\vec B = \frac{Q_{\psi}^{(m)}}{4\pi r^3}\vec r~,
\label{12}
\end{equation}
One may object that in the unitary gauge the  potentials $X_{\mu}$ and ${\cal X}_{\mu}$ are equivalent, and thus no classical magnetic field is possible since the solution ${\cal X}_{\mu}=0$ represents the vacuum. Remarkably enough, the reasoning just given, turns out to be incorrect.  The point is that, in the monopole sector, two potentials ${\cal X}_{\mu}$ and $X_{\mu}$ are gauge equivalent only locally, not globally. Not suprisingly, corresponding field strengths are different. The magnetic field (\ref{12}), corresponding to $X_{\mu}$, we identify as the physical magnetic field of a point-like magnetic charge. 

\paragraph{2.} 
Next we ask the question: Does the Dirac quantization condition hold in the massive theory considered above? The answer is yes\footnote{The authors of  ref. \cite{Ignatiev:1995qz} reached the opposite conclusion. However, they have considered gauge non-invariant Proca theory. We stress again, that gauge invariance (spontaneously broken though) is vital in our consideration. }.  This must be clear as we recall that due to the gauge invariance, the monopole-providing gauge potential $X_{\mu}$ (and not ${\cal X}_{\mu}$) couples to a current density, just like in the ordinary massless theory. That is to say, when we place an electric charge in the field of an "elementary" monopole of charge $g_m$,  the Dirac string feels the electric charge (density) $Q_{\psi}^{(e)}j^{0}$, not the full screened charge,  $Q_{\psi}^{(e)}j^{0}-M_X^2{\cal X}^0$. Consequently, we have 
\begin{equation}
Q_{\psi}^{(e)}g_m=\frac{n}{2}~,
\label{13}
\end{equation}
$n$ is an integer. This is to say that possible minimal ("elementary") electric charge is $g_e=1/2g_m$. Furthermore, it is obvious that the generalization of the Dirac quantization condition to the  case of two dyons $\psi$ and $\xi$ with charges $(Q_{\psi}^{(e)}, Q_{\psi}^{(m)})$ and 
$(Q_{\xi}^{(e)}, Q_{\xi}^{(m)})$, respectively \cite{Schwinger:1966nj}, \cite{Zwanziger:1969by}, 
\begin{equation}
Q_{\psi}^{(e)}Q_{\xi}^{(m)}-Q_{\psi}^{(m)}Q_{\xi}^{(e)}=\frac{n}{2}~,
\label{15}
\end{equation} 
also holds in the massive theory.   If we apply (\ref{15}) to the dyon $\psi$ and its CP-conjugated dyon $\psi^{\rm CP}$ with elementary magnetic charge, i.e. $Q_{\psi}^{(m)}=Q_{\psi^{\rm CP}}^{(m)}=g_m=1/2g_e$ we obtain that dyonic charge is either integer or half-integer of $g_e$, 
\begin{equation}
Q_{\psi}^{(e)}=ng_e~~{\rm or}~~Q_{\psi}^{(e)}=\left(n+\frac{1}{2}\right)g_e~.
\label{16}
\end{equation} 
Quantization of the screened charges has been previously obtained in \cite{Dvali:2007hz} from arguments based on black hole physics. In that case, the charges, although unobservable locally, must still be observable in the Aharonov-Bohm type of interference experiments. In our case, the quantization applies also to the case with no remnant discrete $Z_N$ charges, i.e. when $U(1)_X$ is fully broken, $U(1)_X\to 1$.   

At this point we pause to stress the analogy between the above charge quantization phenomenon and the Aharonov-Bohm effect which is known to persist also in the massive electrodynamics \cite{Krauss:1988zc} (see generalizations in \cite{Dvali:2006az}). The key physical reason is indeed similar: due to the underlying gauge invariance, only unscreened charge does participate in topological Aharonov-Bohm interactions with the $U(1)_X$ string.

In the case of massless $U(1)_X$, the charge quantization (\ref{16}) is strictly correct only in CP-conserving theories. It is well known \cite{Witten:1979ey} (see also \cite{Kobakhidze:2008em}) that in the presence of CP violation the electric charge dequantize. Since the magnetic charge is unscreened, the dequantization phenomenon must hold also in the massive case,  
\begin{equation}
Q_{\psi}^{(e)}=\left(n+\delta \right)g_e~,
\label{17}
\end{equation}
where $\delta$ is non-integral contribution proportional to the strength of CP violation. 

\paragraph{3.} 

Now let us assume that some of the known matter particles constituting macroscopic bodies carry magnetic charges under the massive $U(1)_X$. If the total magnetic charge of a macroscopic body is non-zero, then such charges are severally constrained from the "fifth-forth" experiments.  Suppose, for definiteness, that the electron has an "elementary" magnetic charge $g_m$. Then from the considerations similar to those leading to the bound on the local leptonic charge \cite{Lee:1955vk}, \cite{Okun:1969ey}, we obtain, 
\begin{equation}
g_m\lesssim 4\cdot 10^{-25}~.
\label{18}
\end{equation}  
Thus the magnetic charge must be tiny. However, due to the quantization (\ref{13}) the above bound implies that the minimal electric charge must be enormous, 
\begin{equation}
g_e\gtrsim 10^{24}~. 
\label{19}
\end{equation}
Of course, such a theory is far away from the perturbative domain, and we can not say much about its quantum-mechanical validity. Even the assumed spontaneous symmetry  breaking might be questioned. 

Let us ignore these theoretical troubles here and take more phenomenological attitude. Consider then the Aharonov-Bohm experiment \cite{Aharonov:1959fk} with single solenoid carrying  the ordinary magnetic field $B_e$. The current in the solenoid will produce the $U(1)_X$ magnetic field which due to the Meissner effect is expected to be confined near the surface of the solenoid. However, the magnitude of this magnetic field $B=\frac{g_e}{e}B_e$ is enormous, e.g. $B\approx 10^{24}$T$\approx 10^{8}$GeV$^2$ for a solenoid with $B_e= 1$T.  This value of $B$ is larger than the critical value $B_{\rm crit}\sim \frac{M_H^2}{g_e}\approx 10^{-24}M_H^2$ (see, e.g. \cite{Adelberger:2003qx}), unless the Higgs boson is extremely heavy, $M_H > 10^{16}$GeV (the grand unification scale !). If $B > B_{\rm crit}$, then the magnetic field penetrates into the solenoid through the crietion of Abrikosov  vortices, and the theory inside the solenoid is effectively massless. Then, one must observe additional phase shift related with extra $U(1)_X$ in the Aharonov-Bohm interference experiments.  

In the CP-violating case the minimal $U(1)_X$ electric charge carried by electron is the induced charge $Q_e^{(e)}=\delta g_e$. Thus electron-$U(1)_X$ gauge boson interaction can be made perturbative if CP violation is taken to be small $\delta \lesssim 10^{-24}$. Magnetic field inside the solenoid is in the Proca regime now, and thus the Aharonov-Bohm interactions are also screened. One may also consider a mixing of massive $U(1)_X$ with the ordinary photon through the CP-violating Chern-Simons term, $\frac{\theta g_e e}{8\pi^2}F_{\mu\nu}\tilde X^{\mu\nu}$. This term will induce shifts in ordinary electric charges of particles (providing they also carry $U(1)_X$ magnetic charges). In general, these shifts will cause violation of strict electric neutrality of atoms, which is severely constrained by observations.           
  
\paragraph{4.}
To conclude, we have argued in this paper that magnetic monopole in spontaneously broken $U(1)$ theories produces long-range magnetic force. We have also shown that the Dirac quantization condition for locally unobservable charges holds if CP is an exact symmetry, and the charges are dequantized if CP is broken. Non-observation of the "fifth-force" severely constraints possible $U(1)_X$ magnetic charges of ordinary matter particles (see (\ref{18})). At the same time, the these bounds imply that possible $U(1)_X$ electric charges might be extremely large (\ref{19}), so that their effects in principle can be observed in Aharonov-Bohm  tabletop experiments even if the scale of $U(1)_X$ symmetry breaking is as high as $\sim 10^{16}$GeV. Some other interesting consequences following from the findings in this paper will be discussed in a separate publication.

\subparagraph{Acknowledgments.} I am grateful to Robert Foot and Tony Gherghetta for discussions. 
The work was supported by the Australian Research Council.



\end{document}